\documentclass[conference]{IEEEtran}

\usepackage{graphicx}
\usepackage{url}

\usepackage{breakurl}
\usepackage[breaklinks]{hyperref}
\usepackage{adjustbox}
\usepackage{threeparttable}
\usepackage{float}

\begin{document}

\title{Exploring Widevine for Fun and Profit}

\author{\IEEEauthorblockN{Gwendal Patat}
\IEEEauthorblockA{\textit{Univ Rennes, IRISA, CNRS} \\
Rennes, France \\
gwendal.patat@irisa.fr}
\and
\IEEEauthorblockN{Mohamed Sabt}
\IEEEauthorblockA{\textit{Univ Rennes, IRISA, CNRS} \\
Rennes, France \\
mohamed.sabt@irisa.fr}
\and
\IEEEauthorblockN{Pierre-Alain Fouque}
\IEEEauthorblockA{\textit{Univ Rennes, IRISA, CNRS} \\
Rennes, France \\
pierre-alain.fouque@irisa.fr}
}

\thispagestyle{plain}
\pagestyle{plain}

\maketitle

\begin{abstract}
For years, Digital Right Management (DRM) systems have been used as the go-to solution for media content protection against piracy. With the growing consumption of content using Over-the-Top platforms, such as Netflix or Prime Video, DRMs have been deployed on numerous devices considered as potential hostile environments. In this paper, we focus on the most widespread solution, the closed-source Widevine DRM.

Installed on billions of devices, Widevine relies on cryptographic operations to protect content. Our work presents a study of Widevine internals on Android, mapping its distinct components and bringing out its different cryptographic keys involved in content decryption. We provide a structural view of Widevine as a protocol with its complete key ladder. Based on our insights, we develop WideXtractor, a tool based on Frida to trace Widevine function calls and intercept messages for inspection. Using this tool, we analyze Netflix usage of Widevine as a proof-of-concept, and raised privacy concerns on user-tracking. In addition, we leverage our knowledge to bypass the obfuscation of Android Widevine software-only version, namely L3, and recover its Root-of-Trust. 
\end{abstract}

\section{Introduction} \label{sec:introduction}

Nowadays, people prefer media consumption on over-the-top platforms (OTT), such as Netflix and Amazon Prime, that distribute multimedia content over the Internet, allowing users to play them whenever they wish. Such ease of viewing the same videos across devices creates challenges for content producers and owners. The main challenge remains piracy; namely, platforms delivering media content would like to ensure that the receiving devices offer enough security at the level of hardware and software to prevent leakages. Indeed, protection by authentication is not enough, as the OTT platforms need to prevent free redistribution of copyright protected content.    

The \textit{de facto} answer to these challenges is Digital Rights Management. DRM is a technology that is designed to prevent piracy of digital content. It protects the content owners by restricting media consumption to authorized consumers. Despite begin cried out, DRM systems got increasingly adopted on users' devices. The World Wide Web Consortium (W3C) has recently published the Encrypted Media Extensions (EME)~\cite{EME}, which is the first official Web standard for DRM, ignoring all the expressed worries~\cite{DBLP:conf/space/Halpin17} about a web that should remain open. One of the most popular DRM solutions is Google's Widevine~\cite{widevineSite}, which is currently deployed on web browsers (eg., Chrome and Firefox), Android OS on mobile devices and smart TVs among others. Most popular OTT players and video-streaming services, including Netflix and Disney+, leverage Widevine to protect their content.

Widevine protects video streams at several levels. At the heart of its protection is CENC (Common Encryption Protection Scheme)~\cite{CENC}, specifying encryption standards and key mapping methods that a DRM content decryption module (CDM) should implement to decrypt media files. Nevertheless, the actual key exchanges and protection mechanisms are not documented, because of the proprietary nature of Widevine.

\subsection{Motivation}
Despite the widespread of Widevine, surprisingly, not much attention has been given to its underlying protocol design and security. The main reason behind such a lack of public security analysis is that the DMCA's 1201 clause makes it illegal to study DRM systems. The result is that, under the DMCA, researchers cannot investigate security vulnerabilities if doing so requires reverse engineering. This law has already been used against security researchers to censor their work, as shown by Hewlett-Packard against Snosoft in 2002~\cite{Snosoft}. Unfortunately, more cases have followed, and over fifty court-cases have been launched against research as of 2016~\cite{reported_case_list}.

Fortunately, restrictions have been partially lifted recently. Indeed, in October 2018, the American Library of Congress and the Copyright Office have expanded the exemptions to the DMCA's 1201 clause. Consequently, in theory, security researchers can now freely investigate, correct and publish security flaws on DRM solutions. However, such exemptions did not stop Google that, in November 2020, took down all Github repositories including secret keys of Widevine. That does not mean that exploits compromising Widevine have never been published before. Indeed, the MITRE CVE database~\cite{cve_mitre} lists 25 CVE records since 2014, explaining different security issues within Widevine implementations. Despite such a public record, there is not much literature providing deep insights about Widevine security.

\subsection{Our Contributions}
Our work aims at pushing this topic forward, as we believe that stakes are high regarding the Widevine protocol. Our paper intends to fill this lack of public research by providing the first thorough analysis of the Widevine protocol. Here, we overcome the restriction of signing a non-disclosure agreement (NDA) to get the full description of the Widevine protocol by performing a complete reverse engineering of the Android Widevine modules. Moreover, we show that a deep understanding of Widevine can allow attackers to easily recover its internal secret parts without requiring to defeat the applied obfuscation. This is worrisome for two reasons. First, many streaming services rely on Widevine DRM to protect content against piracy. Any harm to this technology can lead to huge financial losses. For instance, in 2020, the OTT market size was estimated at \$13.9 billion and expected to reach \$139 billion by 2028~\cite{ott_worth}. Second, zero-day vulnerabilities on Widevine can be exploited to harm countless users~\cite{laginimainebWidevine}, since Widevine is estimated to be installed on more than 5 billion devices around the world.

In this work, our approach was to begin with a manual analysis to gain insights into the structure of the Widevine protocol as well as its main cryptographic operations. Then, we design a Frida-based tool to automatically extract details about the Widevine workflow as it is leveraged by Android OTT apps. Afterwards, we extend our tool to trace the Widevine operations within web applications running in an EME-supporting browser. Finally, we take a look at the outcome of our analysis, and discuss its relevance regarding the security of the Widevine module or the OTT app. We emphasize that, throughout our work, we carefully play the role of the security researchers who, as described by the DMCA's exemptions, act in good faith. Indeed, we timely report all our findings to Google, Widevine and Netflix. In addition, we gave up all the keys that we succeeded to extract, so that they get revoked by the concerned parties. \\

\noindent
Our contributions are the following:
\begin{itemize}	
	\item We reverse-engineer Widevine components on Android. In particular, we thoroughly explore its different cryptographic components from the root of trust, aka \textit{keybox}, until the key decrypting the media, aka \textit{Content Key}, and provide an implementation of this key ladder.\footnote{\url{https://github.com/Avalonswanderer/widevine_key_ladder}}
	
	\item We uncover the structure of the Widevine protocol and detail the contents of the exchanged messages between Widevine and the different entities of the DRM ecosystem. We also dissect Widevine internals during the execution of an OTT app, and split them into three main operations: provisioning of device-specific RSA key, provisioning of content license keys, and content decryption.

	\item In order to automate Widevine inspection, we design \textit{WideXtractor},\footnote{\url{https://github.com/Avalonswanderer/wideXtractor}} which is a tool based on Frida to monitor the inner working of Widevine during media playback and dump exchange messages. 	
	
	\item We leverage our tool to study Netflix usage of Widevine, in order to understand how it protects its media assets: video, audio and subtitles. We find that Netflix mainly establishes two Widevine sessions: one to receive protected media, and another one for obtaining the related decryption keys. We spy on the first session using WideXtractor, and notice that Netflix, unlike other OTT apps, only protects the download URL of audio tracks, that can be effortlessly downloaded in clear even without a Netflix account. Indeed, the second session only concerns the decryption key of video tracks.
	
	\item We discuss two issues raised by our analysis. The first one is related to the use of a distinctive device identifier by Widevine. This allows third-party servers to profile users' behavior during media consumption without their consent. The second one concerns a methodology that we define to efficiently recover Widevine Android software-only root of trust despite the underlying obfuscation hiding the critical parts of Widevine. We timely report all findings to Google and Netflix and were assigned the CVE-2021-0639. We were awarded by the bug bounty program of Netflix and the Vulnerability Reward Program of Google.

\end{itemize}

\noindent
\textbf{Roadmap.} This paper is organized as follows: \autoref{sec:background} introduces the necessary background for our research. Section~\ref{sec:android_drm} offers an overview of the Android DRM and Widevine plugin components interaction. The Widevine protocol is presented in \autoref{sec:widevine_internals} with a special focus on its key ladder in \autoref{sec:crypto_ladder}. Our reverse engineering efforts allow us to implement WideXtractor that automatically inspects Widevine flow for OTT apps. We present the design of our tool in \autoref{sec:widevine_tool} and provide some insights gained by applying it on several OTT apps. Section~\ref{sec:discussion} discusses the concerns and security issues that emerge from our study. Related works are presented in \autoref{sec:related_work}. We conclude in \autoref{sec:conclusion}.

\section{Background} \label{sec:background}

\subsection{Digital Right Management Systems} 
Digital Rights Management (DRM) system refers to technology that plays two roles. First, it offers the tools that enable a content provider to encrypt their content. Second, it builds an ecosystem, so that the content provider controls who can decrypt and consume their content. To this end, DRM systems define a set of business rules to be enforced. In practice, DRM involves two parties: a Content Delivery Network (CDN) supplying encrypted content and a License Server providing the necessary keys to decrypt such content. Only the DRM module on the user device can retrieve these keys, which makes it possible to control media consumption. The DRM module shall also protect the keys while using them. For instance, Widevine relies on ARM TrustZone based Trusted Execution Environment (TEE) when available for strong security guarantee in many Android devices.

The DRM module performing sensitive operations, such as decryption and license requests, is implemented separately and called CDM or Content Decryption Module. Every DRM scheme provides its own CDM that includes proprietary mechanisms for License Server communication, as well as rules around local license storage and renewal. Of course, CDMs are required not to leak license keys. By being closed-source, CDMs mostly rely on security-by-obscurity.

\subsection{Google Widevine}
Widevine is a DRM solution acquired by Google in 2010. The earlier version of Widevine that had support in old Android versions (up to Android 5.1) was called Widevine Classic, working only with the proprietary \textit{.wmv} format. The current version of Widevine is called Widevine Modular, and implements a different DRM and streaming standards, including MPEG-DASH and CENC. Widevine Modular, or henceforth simply Widevine, is supported on Android 4.4+.

Widevine defines three security levels: L1, L2 and L3, where the L1 level is considered the most secure for playing HD videos from OTT platforms. Widevine depends on the TEE to implement L1 security. At L1, both cryptography and video processing take place inside the TEE. It is worth noting that applications inside the TEE are hardware dependent, and therefore Widevine shall provide a different implementation for each one. L2 and L3 are implemented where the TEE is not an option, such as legacy phones or Widevine locked ones due to device tampering. In Android, Widevine does not propose L2 security. The L3 lets both cryptography and video processing take place outside the TEE. They are considered more vulnerable, given that the CDM is software-only. As for users, L3 delivers sub-HD resolutions since Widevine supports HD and ultra-HD content only for L1.

\section{Warm Up: Widevine in Android} \label{sec:android_drm}

In this section, we describe the integration of Widevine into the Android ecosystem. In particular, we detail all the components of which Widevine consists of and their interaction. This will help us to better frame our reverse engineering methodology by pointing out the relevant components to analyze in order to uncover the internals of  Widevine.

\subsection{Android DRM API}
In order to cope with the fragmented DRM ecosystem, Android offers a unified API in Java/Kotlin for DRM systems. Starting from API level 18, this is implemented by some HAL (Hardware Abstraction Layer) module called Media DRM Server that abstracts the actual running DRM from the programming interface used by OTT apps. The Android DRM API mainly consists of two modules: Media DRM and Media Crypto. The Media DRM is used to communicate with License Servers and to manage keys for a given media. As for Media Crypto, it is used to perform decryption. The DRM APIs support the ISO/IEC 23002-7: Common Encryption standard (CENC)~\cite{CENC}, but implement other encryption schemes. 

Playing encrypted content when leveraging DASH (Dynamic Adaptive Streaming over HTTP) works as follows. First, the app constructs a Media DRM object with a given DRM through a unique identifier. Then, the app opens a new session with the Media DRM object and gets some session identifier. A Media Crypto object is then constructed and bound to the opened session. Next, Media DRM retrieves keys (aka licenses) from the License Server. To this end, a DRM-specific request object is obtained from the Media DRM object, and the server response is delivered to the Media DRM instance. The obtained keys are only accessible through Media Crypto. Indeed, the encrypted content is decrypted by a Media Codec instance to which the Media Crypto object was registered. Thus, the keys are not accessed directly. In addition to the DASH mode, the DRM APIs provide the ability to establish a secure session to protect arbitrary data.

\subsection{Widevine Components}
In Android, Widevine comes as a dynamically loadable HAL plugin within the \texttt{mediadrmserver} process. Similar to other HAL plugins, Widevine is manufacturers-provided. In addition, it is not open-source; only provided as binary code and library files. To keep things secure, when a TEE is available, the HAL plugin delegates all sensitive operations to the Widevine component that runs inside the TEE. Roughly speaking, the resulting architecture looks like this (other components might exist depending on the Android version):
\begin{itemize}
	\item \textbf{Widevine library:} this library is used by the \texttt{mediadrmserver} process to translate Android DRM API calls to Widevine CDM ones. The behavior of this library changes depending on the Widevine security level. In L1, it plays the role of a proxy and communicates with the TEE through \texttt{liboemcrypto.so}. As for L3, it contains the obfuscated CDM. Its name can change depending on the version and SoC including but not limited to: \texttt{libwvdrmengine.so}, \texttt{libwvhidl.so}, \texttt{libwvm.so}, \texttt{libdrmwvmplugin.so}.
	\item \textbf{liboemcrypto.so:} this library performs marshalling and unmarshalling of requests to the Widevine trustlet. All communications with the TEE go through a specific TEE driver (e.g., \texttt{QSEEComAPI.so} for QSEE).
	\item \textbf{Widevine trustlet:} it runs inside the TEE and implements all the needed functionalities for L1.
\end{itemize}

\subsection{Components Interaction}
\begin{figure}
	\centering
	\includegraphics[scale=0.5]{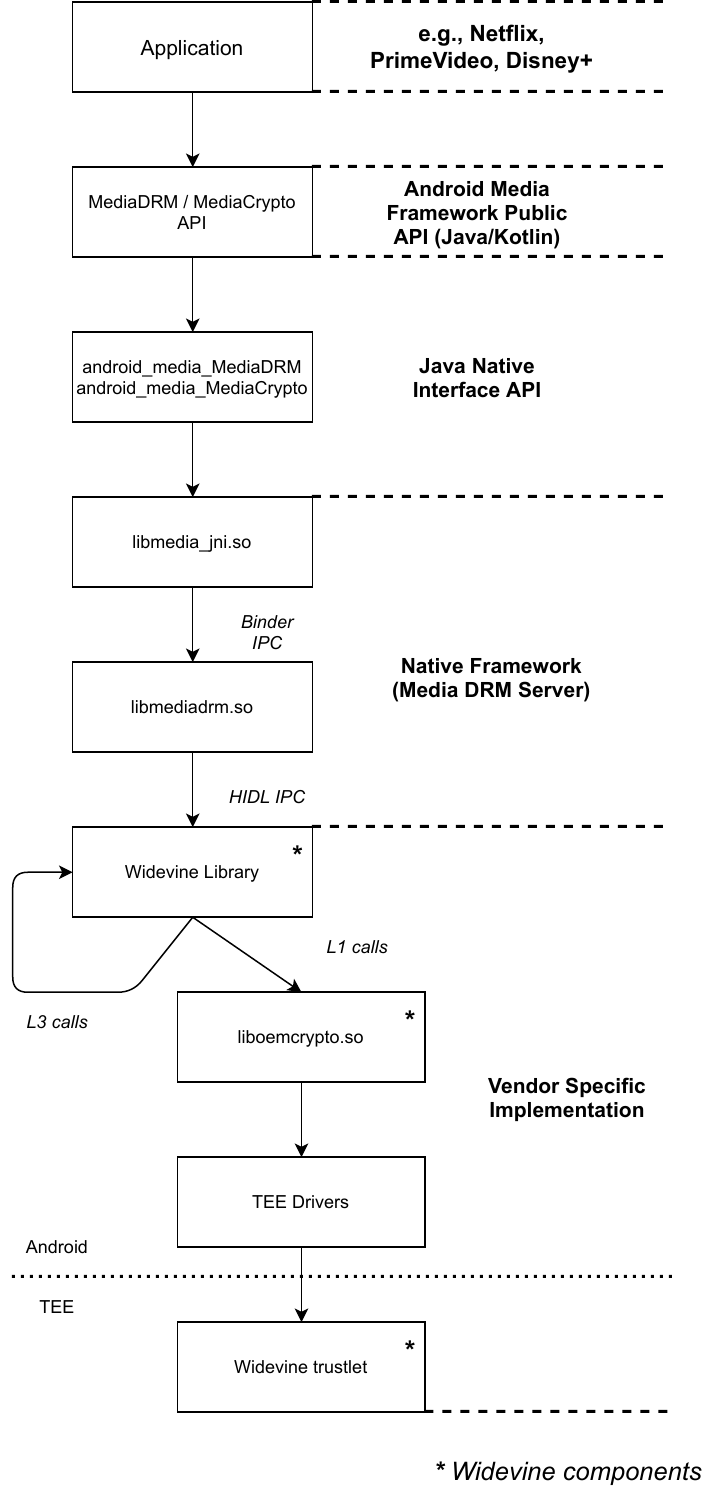}
	\caption{Widevine Architecture in Android}
	\label{fig:components}
\end{figure}

Android Widevine architecture is summarized in \autoref{fig:components}. In a top down architecture, components interact as follows. DRM services start from the OTT application calling the Android Media Framework API to interact with the \texttt{MediaDRM} and \texttt{MediaCrypto} objects. All calls to the DRM API go through some Java Native Interface (JNI) layer via the \texttt{libmedia\_jni.so} library. Calls are then forwarded to the Media DRM Server instantiated by the \texttt{mediadrmserver} process, which is the last module implemented by Android. The Media DRM Server reaches the Widevine specific implementation through the HAL APIs. Any communication with Widevine first goes to its specific library such as \texttt{libwvdrmengine.so}. In L3, no further component is involved. As for L1, whenever CDM is required, this library calls \texttt{liboemcrypto.so} that sends the related requests to the Widevine TEE trustlet.

Of particular interest, the Widevine library does the translation between the HAL API to Widevine functions. Once translated, if Widevine is in L1 mode the Widevine API is used to call its equivalent in \texttt{liboemcrypto.so}. The OEMCrypto library role is to forge a message for the TEE trustlet containing the function arguments and command code. For a given TEE, each Widevine function corresponds to a specific command code used by the TEE driver in order to be received by the Widevine command handler within the trustlet.

\section{Widevine Internals} \label{sec:widevine_internals}

\begin{figure*}
	\centering
	\includegraphics[scale=0.48]{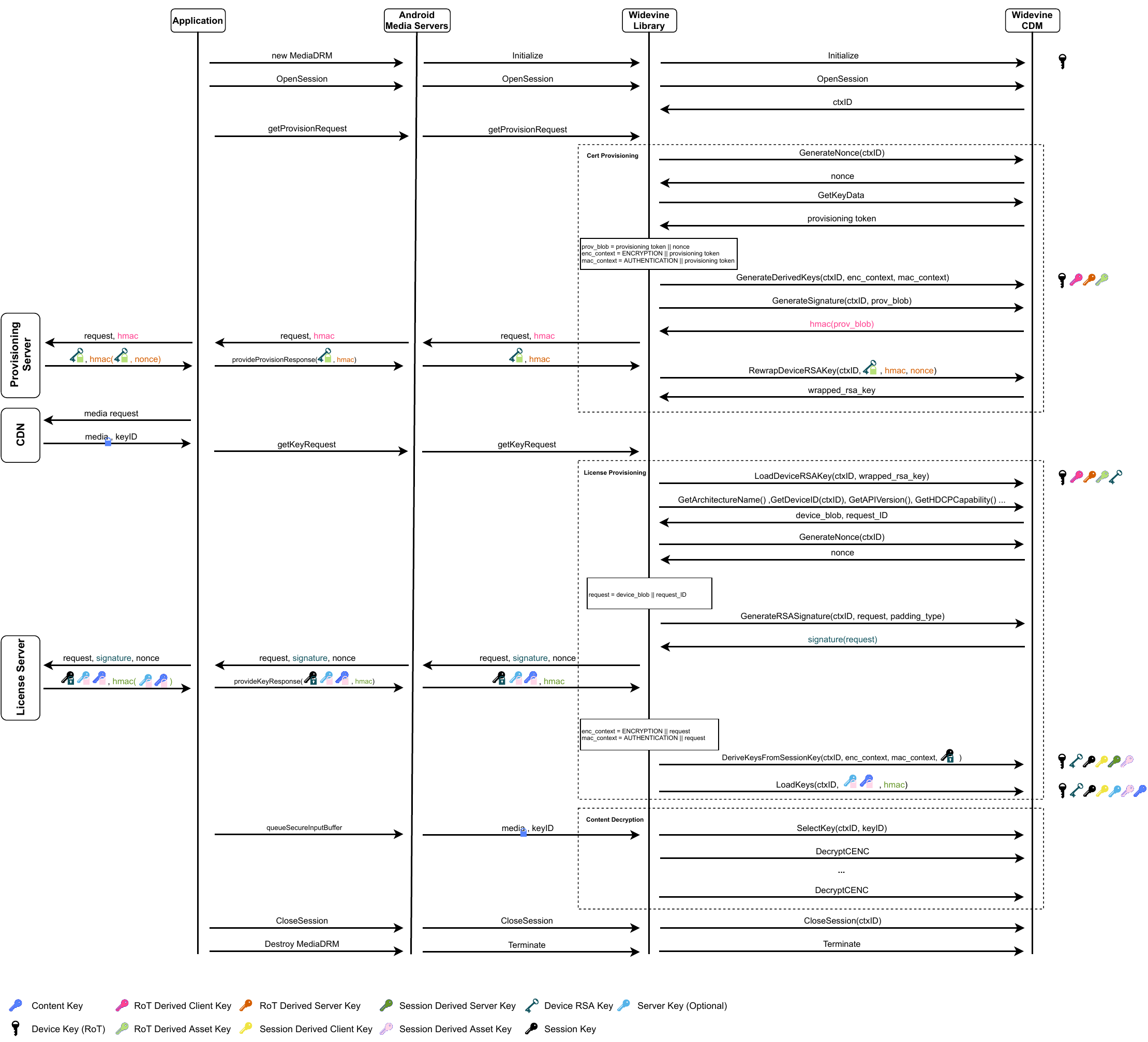}
	\caption{Widevine Protocol for Media Decryption}
	\label{fig:protocol}
\end{figure*}

\subsection{Methodology}
Given the Widevine component layout described in \autoref{sec:android_drm}, three main components of Widevine come to light: the Widevine library, \texttt{liboemcrypto.so}, and the Widevine trustlet running in the TEE. To study Widevine inner working, we have inspected both statically and dynamically these components starting from the Widevine trustlet.

\textbf{Analysis Environment.}  We examine four smartphones and their respective factory images:

\begin{itemize}
	\item \textbf{Nexus 5}: Android version 6.0.1, build hammerhead m4b30z. Widevine L3 mode with version 3.1.0.
	\item \textbf{Nexus 5X}: Android version 8.1.0, build bullhead bhz32c. Widevine L1 mode, version 5.1.0, security patch level 2. 
	\item \textbf{Pixel:} Android version 10, build sailfish qp1a.191005.007.a3. Widevine L1 mode version 14.
	\item \textbf{Pixel 3:} Android version 11, build blueline rq3a.210805. Widevine L1 mode version 15. 
\end{itemize}
All tested smartphones integrate the Qualcomm TEE (QSEE), and therefore our analysis includes some QSEE-related details. Henceforth, we will be careful to distinguish what is true for Widevine as a protocol, and what is specific to QSEE.

\textbf{Trustlet Extraction.} The first step to reverse the proprietary implementation of Widevine was to extract the binaries from the file system either from a physical (rooted) mobile or by downloading and extracting the phone factory images on the Google website~\cite{googleFactoryImages}). These binaries, including the trustlet, can be found in the \texttt{/vendor/} directory.

For Qualcomm SoC based phones, Widevine trustlet is located in \texttt{/vendor/firmware/} and is divided into several .bXX files, where XX is a counter. To reconstruct the complete trustlet, we can simply concatenate each of these files in order to obtain a standard ELF binary with an additional hash table section used for integrity verification.

\textbf{Trustlet Analysis.} The Widevine L3 CDM within the Widevine library (e.g., \texttt{libwvdrmengine.so}) being obfuscated, we preferred to start with another implementation in order to figure out how Widevine works. We notice that the L1 trustlet is not obfuscated at all. Worse still, it provides verbose clear debugging strings in its internal code, thereby leading to a better understanding of the overall structure and control flow of the protocol. Leveraging tools such as \textit{Ghidra}~\cite{Ghidra} and \textit{Radare2}~\cite{Radare2}, we were able to retrieve function names and cryptographic keys within the Widevine protocol. Relying on our knowledge of the L1 trustlet, we then went back to the library to discern the HAL calls to L1 and to analyze the obfuscated L3 layer.

\textbf{Libraries Analysis.} Our reverse engineering of the trustlet allowed us to map the Widevine protocol functions called OEMCrypto with function symbols (\textit{\_oeccXX} for L1 and \textit{\_lccXX} for L3). The correspondence for Widevine Modular L1 functions is summarized in~\autoref{appendix:oeccSymbol}. We also leveraged the \textit{Frida} toolkit~\cite{Frida} to trace the execution of these functions in the \texttt{mediadrmserver} process, as both the Widevine library and \texttt{liboemcrypto.so} are instantiated in the Media DRM Server. This leads us to figure out the workflow of the Widevine protocol regarding the called operations as well as the related cryptographic keys.

\subsection{Widevine Protocol} 
In \autoref{sec:android_drm}, we present the components enabling protected content playback within Android devices using Widevine. Taking a look at the bigger picture to highlight the communications behind these elements, we can distinguish 7 agents: a CDN (Content Delivery Network), a provisioning server, a license server, an OTT application, the Android Media servers, the Widevine library and Widevine CDM. 

Indeed, when playing protected content using an OTT app (e.g., Netflix), content decryption is managed by the Android Media Servers that relies on the underlying DRM system, here Widevine. All key requests for provision and license servers are generated by Widevine components, especially the library, with the help of the CDM. Overall, the Widevine protocol involving these actors is divided into three main phases: Certificate Provisioning, License Provisioning and Content Decryption. An illustration of the protocol can be seen in~\autoref{fig:protocol}.

\textbf{Certificate Provisioning.} The provisioning phase is usually done once to recover a cryptographic certificate and does not need to be done for future media decryption. The private key within this certificate protects the fresh session keys. A new request is sent to the provisioning server when no certificate can be found, the one installed is corrupted, or the OTT needs to install a new certificate.

On request creation, the CDM generates a nonce to ensure freshness. Then, it derives keys for certificate decryption and integrity checks, based on the Widevine Root of Trust (RoT) called the \textit{Device Key}, and dynamically generated buffers. These buffers are based on a token within the RoT structure detailed later. Using \texttt{OEMCrypto\_GenerateSignature}, the request is HMAC-protected with the RoT derived client key, and sent to the provisioning server.

The received response is passed to the CDM through the \texttt{OEMCrypto\_RewrapDeviceRSAKey} function. After nonce check and integrity verification using the RoT derived server key, the certificate is decrypted using the previously derived key and stored on the persistent storage of the device after being rewrapped (i.e. re-encrypted) by a device-specific key. This marks the end of the installation process of the certificate private key called the \textit{Device RSA Key}.

\textbf{License Provisioning.} The OTT receives all required information about the protected content from the CDN in order to ask for the corresponding \textit{Content Keys}, also known as license keys. After loading the stored certificate using \texttt{OEMCrypto\_LoadDeviceRSAKey}, a request is forged using a generated request ID and various device specific info concatenated within a device blob. This message is then signed by the Device RSA Key and sent with a newly generated nonce to the License Server.

The corresponding response will be received by the CDM through \texttt{OEMCrypto\_DeriveKeyFromSessionKey} that will extract and decrypt a \textit{Session Key} using the \textit{Device RSA Key}. The \textit{Session Key} is then used to derive other HMAC and encryption keys based on buffers containing a dynamically generated device blob. These keys are later used to verify the integrity of the response received by \texttt{OEMCrypto\_LoadKeys} and to decrypt the \textit{Content Keys}. Here, Content Keys are associated with a Key Control Block (KCB), encrypted or not by its related content key. KCB contains the previous nonce and various information that we will detail in \autoref{subsec:contentkey}. It is important to note that, once all Content Keys have been added to the CDM, the nonce is cleared from memory. During key reception, the License Server can provide a new Server Key protected by the derived asset key. This key constitutes a new server mac key for integrity verification of future response, such as in \texttt{OEMCrypto\_RefreshKeys}. 

\textbf{Content Decryption.} Note that more than one Content Keys can be loaded in the CDM memory at the same time. Therefore, the right one for the media is selected using its key ID in \texttt{OEMCrypto\_SelectKey} before content decryption within \texttt{OEMCrypto\_DecryptCENC}.

\section{Widevine Crypto Ladder} \label{sec:crypto_ladder}

\begin{figure}
	\centering
	\includegraphics[scale=0.55]{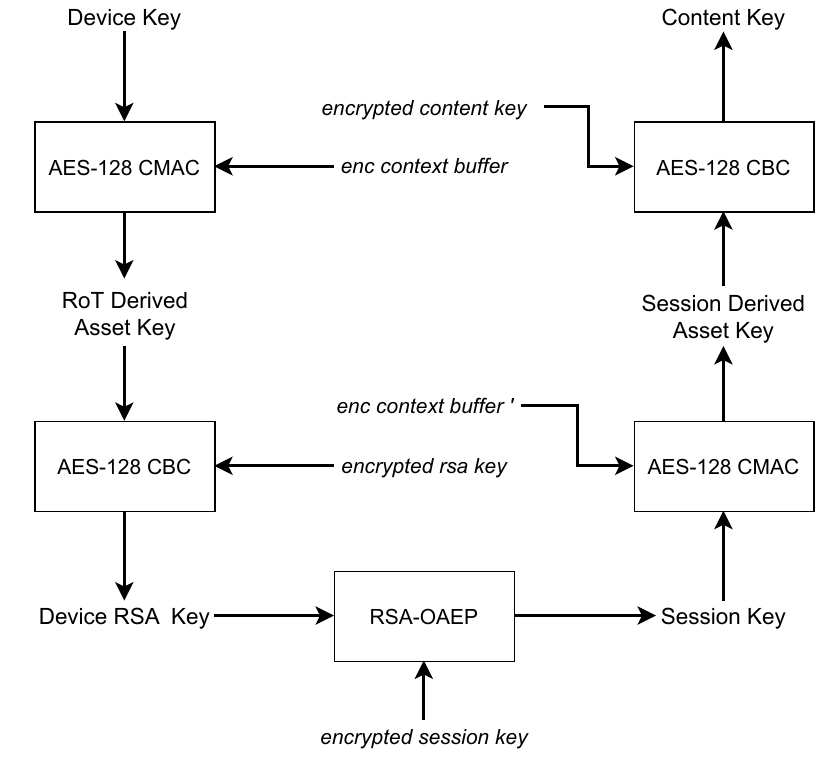}
	\caption{Widevine Key Ladder}
	\label{fig:key_ladder}
\end{figure}

In this section, following our reverse engineering analyses, we succeed in depicting a complete picture of Widevine cryptographic mechanisms. In particular, we uncover the Widevine internal key ladder from its root of trust to the content decryption key. An overview is summarized in \autoref{fig:key_ladder}. Note that our study holds for both L1 and L3. Moreover, we provide an implementation of the key ladder in Python.\footnote{\url{https://github.com/Avalonswanderer/widevine_key_ladder}} For ethical and legal purposes, we did not include the root keys. Thus, nobody can use our project to actually pirate OTT contents.

\subsection{Widevine Root of Trust: Keybox} \label{subsec:keybox}
Widevine, in~\cite{WidevineSiteKeybox}, mentions that its RoT is established by a factory-provisioned component called the \textit{keybox}. While doing our static analysis, we noticed that this keybox is mainly used to secure the provisioning of certificates. We will discuss more about this in~\autoref{subsec:cert_prov}. Now, let us focus on the nature of this RoT and where it is stored.

\textbf{Keybox storage.}
While exploring the structure of the Widevine trustlet, we looked at the Qualcomm Secure File System (SFS) that allows trustlets to encrypt their sensitive data before storage. On our setup, the SFS persistently stores the encrypted files on \texttt{/persist/data/sfs/}. Then, we continued our exploration and found the function \texttt{init\_tzdrm\_config\_path} initializing paths for several elements. Here, the string names are explicit, and we think that the \textit{keybox\_lvl1.dat} file refers to the keybox for L1. We also found the \texttt{OEMCrypto\_InstallKeybox} function that can be called from \texttt{libwvdrmengine.so} to re-install a factory keybox supposed to exist in the \texttt{/factory/wv.keys} file. Nevertheless, we have never found such a file. As for L3, we noticed that the library \texttt{libwvdrmengine.so} loads the keybox from the \texttt{ay64.dat} file that can be found in \texttt{/data/mediadrm/IDMXXXX/L3/}.

\textbf{Keybox structure.}
In order to figure out of what the keybox consists, we looked deeper at the trustlet functions that verify keybox integrity, because they display an explicit error log message that clearly identifies the concerned field. We began with \texttt{OEMCrypto\_isKeyBoxValid}. Indeed, this function shows us that the keybox structure is 128 bytes long with two special fields at the end. The trustlet checks the integrity of the decrypted keybox by verifying that the last eight bytes are composed of a magic number \textit{``kbox''} followed by a 4-byte Cyclic Redundancy Check (CRC-32) code. In this paper, we will not discuss the effectiveness of this integrity verification despite being broken in other contexts~\cite{DBLP:conf/uss/Garman0KMR16}. We retrieved the remaining fields by looking at the functions loading the keybox: \texttt{OEMCrypto\_GetKeyData}, \texttt{OEMCrypto\_GetDeviceID}, and the internal function \texttt{OEMCrypto\_GetDeviceKey}. The \textit{Device ID} is a 32-byte unique device identifier for Widevine. The \textit{Key Data}, or the \textit{Provisioning Token} as referred to by \texttt{libwvdrmengine.so}, is 72 bytes long and is used wihtin provisioning requests. The remaining 16 bytes correspond to an AES key that is called, according to the API, \textit{Device Key}. This key is the real RoT. We summarize the five fields of the keybox in~\autoref{tab:widevinekeybox}. 
 
\begin{table}
	\centering
	\caption{Widevine Keybox}
	\resizebox{0.48\textwidth}{!}{
		\begin{tabular}{|c|c|c|}
			\hline
			\textbf{Field} & \textbf{Description} & \textbf{Size (bits)} \\ \hline
			Device ID & Internal Device ID & 256\\ \hline
			Device Key & 128-bit AES key & 128\\ \hline
			Provisioning Token & Used by provision requests & 576\\ \hline
			Magic Number & ``kbox'' & 32\\ \hline
			CRC32 & CRC32 validating the keybox integrity & 32\\ \hline
			\textbf{Total} &  & \textbf{1024} \\ \hline
		\end{tabular}
	}
	\label{tab:widevinekeybox}
\end{table}

\subsection{Device RSA Key}
\label{subsec:cert_prov}
As explained in \autoref{sec:widevine_internals}, Widevine does not directly use its RoT to protect licenses. Instead, it leverages an RSA key pair that, unlike the keybox, can be installed dynamically on the device through a process called certificate provisioning. This process is protected by the \textit{Device Key} that is derived into several keys in \texttt{OEMCrypto\_GenerateDerivedKeys}. Upon reception, the \texttt{OEMCrypto\_RewrapDeviceRSAKey} function first verifies the integrity of the key pair by re-computing an 256-bit HMAC tag. Then, it decrypts the key pair, and re-encrypts it again with a device-unique (TEE-specific in L1 or keybox-related in L3) 128-bit AES key. These keys never leave the CDM. This function also checks the key pair format after decryption and before re-encryption. Indeed, the key pair, aka the \textit{Device RSA Key}, is expected to be an RSA certificate with PKCS\#5 padding in PKCS\#8 format as indicated by the \texttt{qsee\_secpkcs8\_parse} and \texttt{get\_len\_with\_pkcs5\_padding} functions. 

Once re-encrypted, aka re-wrapped, the \textit{Device RSA Key} is stored on the standard file system by the Widevine library in a \texttt{cert.bin} file on \texttt{/data/mediadrm/IDMXXXX/}. Widevine distinguishes L1 from L3 by using different directories. This persistent data is later used in future instances of the Widevine CDM to avoid making new provisioning. Here, \texttt{OEMCrypto\_LoadDeviceRSAKey} is used to recover the stored certificate. We note that re-wrap is MAC-and-Encrypt, as it also computes an 256-bit HMAC tag on the key pair. Please note that a new provisioning process is performed whenever the \texttt{cert.bin} file is deleted or corrupted.

\subsection{Content Keys} \label{subsec:contentkey}
\textbf{Keys protection.}
As introduced previously, content keys, or license keys, are obtained from \texttt{provideKeyResponse}. Here, Widevine first calls \texttt{DeriveKeysFromSessionKey} from \texttt{OEMCrypto} to decrypt a special field; \textit{Session Key}, using the \textit{Device RSA key}. This key is later used to derive a 128-bit AES key as well as two 256-bit HMAC keys. Then, these keys are used in \texttt{OEMCrypto\_LoadKeys} to decrypt the license key and verify its integrity. 

\textbf{Key Control Block.}
Alongside the Session and Content Keys, the response from the License Server also contains additional 128-bit metadata called \textit{Key Control Block} (KCB), one for each license key in the response, and is encrypted by its associated \textit{Content Key}. The KCB is identified by the magic number \textit{kctl} or one of the form \textit{kcXX}, where XX is related to Widevine version. The \textit{Content Key} is accepted only when the associated KCB is checked by \texttt{verifyKeyControlBlock}. This function helped us to understand the structure of KCB: a nonce, time to live (TTL) of the key, and 32-bit of control bits. These control bits define usage right (e.g., encryption, MAC tag generation, etc.) and anti-rollback features. During the lifetime of a \textit{Content Key}, the KCB can be updated through the \texttt{OEMCrypto\_RefreshKey} function that, unlike its name might indicate, cannot change the key or usage rights but only its TTL. Such refresh requests work as license ones with the exception that the \texttt{OEMCrypto} function \texttt{GenerateSignature} is used for integrity protection instead of \texttt{GenerateRSASignature}.

\textbf{Keys Usage.}
During loading in the CDM memory, Content Keys are stored in a key table structure with an identifying key ID for \texttt{OEMCrypto\_SelectKey}. For media playback, encrypted buffers are decrypted with the chosen key by \texttt{OEMCrypto\_DecryptCENC} implementing MPEG-CENC.

\subsection{Nonces} 
The Widevine protocol mitigates replay attacks and ensures message freshness using nonces. By \texttt{OEMCrypto\_GenerateNonce}, the Widevine library can ask the CDM to generate up to 20 nonces per second stored in a First In, First Out (FIFO) queue of 16 elements within the CDM memory. These 32-bit nonces are generated using a Pseudo Random Number Generator (PRNG) and used at each request creation and response reception. If the nonce is valid, it is removed from the queue and the message is processed, otherwise the message is dropped. During \textit{Content Keys} loading in \texttt{OEMCrypto\_LoadKeys}, a single nonce can be used in multiple KCBs. In this case, the nonce is only removed once all keys have been processed. 

\subsection{Summary of Cryptographic Algorithms} \label{subsec:crypto_algo_sum}
\textbf{Widevine Generic Crypto API.} In addition to media decryption, the Widevine CDM allows applications to perform arbitrary cryptographic operations within a dedicated session. In Android, using the \texttt{CryptoSession} class of Media DRM, an application can leverage the underlying DRM plugin to protect data using the \texttt{OEMCrypto\_Generic\_XXX} family of \texttt{Encrypt}, \texttt{Decrypt} \texttt{Sign} and \texttt{Verify} functions. Here, each operation must have the appropriate key usage rights.

\textbf{Key Derivation.} Widevine never uses received or stored keys directly. Instead, it relies on key derivation algorithms, implemented in \texttt{OEMCrypto\_GenerateDerivedKeys} and \texttt{OEMCrypto\_DerivedKeyFromSessionKey}, in order to generate three different keys each time: a 128-bit Asset key, a 256-bit MAC Client Key, and a 256-bit MAC Server Key. The leveraged PRF (Pseudo-Random Function) is AES-128-CMAC to generate the required 640 bits. In addition to secret keys, the derivation algorithm uses two buffers, called \textit{encryption context} and \textit{mac context}, that are created based on device-unique information and used respectively for the Asset key and both MAC keys. For each chunk, the device blob is appended to a string that begins with a string counter and the word ``\textit{ENCRYPTION}'' for encryption context or ``\textit{AUTHENTICATION}'' for the mac ones. Only one encryption context with counter ``1'' is needed, while the mac context includes four counters starting from 1.

\textbf{Symmetric Cryptography.} All CDM operations related to key decryption, generic encryption API (i.e. \texttt{Generic\_Encrypt}, \texttt{Generic\_Decrypt}), and media protection are performed using AES 128 bits. Both key decryption (e.g. \texttt{OEMCrypto\_LoadKeys}) and encryption API implement AES in CBC mode, while media decryption relies on \texttt{OEMCrypto\_DecryptCENC} which supports MPEG-CENC (i.e. AES-128-CTR).

\textbf{AES Initialization Vectors.} Within key decryption functions, IVs are being received alongside their respective cyphertext in server responses. On \textit{Device RSA Key} rewrap, IV generation is handled by a PRNG algorithm with \texttt{OEMCrypto\_GetRandom}. For protected content, MPEG-CENC standard is used to deal with IVs.

\textbf{MAC Generation.} The MAC Client Keys and Server Keys respectively protect requests and responses to provisioning and license servers using HMAC-SHA256. The same algorithm is used by the \texttt{OEMCrypto} API \texttt{Generic\_Sign} and \texttt{Generic\_Verify} using the currently loaded Content Key. 

\textbf{RSA Operations.} The \textit{Device RSA Key} includes a 2048-bit private key that is used for both decryption and signature during \textit{Session Key} loading and license request creation. For decryption, this key is used in the RSA-OAEP-SHA1 mode, while RSASSA-PSS-SHA1 and RSASSA-PKCS1-v1\_5 can both be used for requests depending on function argument.

\section{WideXtractor} \label{sec:widevine_tool} 

Most OTT apps, including Netflix, Disney+ and Hulu, apply anti-debugging mechanisms in order to prevent attackers from easily intercepting and tracing calls to Widevine. In addition, our reverse engineering shows that it is quite demanding to untangle the Widevine interface between its different components. This is due to the fact that dissecting the Widevine workflow provides an important insight into its internals. Thus, we implement WideXtractor; a tool tracing the Widevine calls made by an OTT app. In this section, we present an overview of the design and the implementation of this tool. Then, we show the effectiveness of our tool by highlighting interesting findings while analyzing the most downloaded OTT app on Android, namely Netflix. Finally, we extend WideXtractor to inspect Widevine as a plugin on Chrome.  

\subsection{WideXtractor Design}
We design WideXtractor to automatically trace the execution flow of Widevine on Android. Our goal is twofold. First, researchers can easily and systematically study how OTT apps leverage Widevine while displaying protected content. Interesting findings can be revealed by analyzing the actual calls and their parameters, as we demonstrate for Netflix in the following subsection. Second, our insights about the used operations make the tool uncover the secret keys that should exist in the memory at a particular point of time, despite of the applied obfuscation.

We implement WideXtractor using Frida to monitor all calls to the \texttt{OEMCrypto} functions in the Media DRM Server. Monitoring the \texttt{mediadrmserver} process has two advantages: (1) it allows us to bypass anti-debug techniques at the application level, such as SafetyNet~\cite{safetynet}, and (2) both L1 and L3 workflow can be recovered.

Our monitoring traces any call to Widevine functions, hence the underlying protocol outline, while dumping the used arguments, such as buffers for requests and opaque reply data. To this end, WideXtractor relies on a Frida server running on the Android device with \textit{sys\_ptrace} capability. This can be achieved by running as the media group or a system/root privileged user. Our script hooks the \texttt{OEMCrypto} functions after attaching to the \texttt{mediadrmserver} process instantiating DRM libraries.

Once launched and attached, WideXtractor logs every method of the Widevine protocol and received buffers from the Android OS. Our traces correspond to the library symbols that we reverse engineered and summarized in \autoref{appendix:oeccSymbol}. Our tool allows attackers to inspect exchanged messages during key reception and media decryption before opaque requests and responses. WideXtractor can be found on our Github.\footnote{\url{https://github.com/Avalonswanderer/wideXtractor}}

\subsection{Case Study: Netflix} 
Leveraging WideXtractor, we automatically monitor the use of Widevine by the Netflix app, which is the most popular OTT with 200 million subscribers around the world~\cite{netflix_verge}. Our analysis shows a large number of calls to the Widevine Generic Crypto API compared to other OTT apps.

Following our observation, we dig deeper to understand Netflix internals. We find that Netflix requires to manage two Widevine sessions: one to get \textit{Content Keys} and decrypt protected media, and one to exchange data using the Widevine Generic Crypto API. Henceforth, we will call these two sessions \textit{License Session} and \textit{Generic Crypto Session} respectively. Both sessions are initialized in the same way until \texttt{OEMCrypto\_LoadKeys}.

Indeed, the License Session loads the \textit{Content Key} that decrypts the displayed media, while the Generic Crypto Session loads several keys for confidentiality and integrity protection of arbitrary data. We note that \textit{Content Key} can only be accessed through the \texttt{queueSecureInputBuffer} method from the \texttt{MediaCodec} class. Thus, only keys loaded within the Generic Crypto Session can be used to decrypt arbitrary data using the Android \texttt{CryptoSession}. Here, each key has its own usage rights to perform specific cryptographic operations. As explained in \autoref{subsec:crypto_algo_sum}, Widevine associates these functions to the following \texttt{OEMCrypto} ones: \texttt{Generic\_Encrypt} and \texttt{Generic\_Decrypt} for AES operations, and \texttt{Generic\_Sign} with \texttt{Generic\_Verify} for HMAC tag computation and verification.

Our study shows that all communications with the Netflix CDN go through the Generic Crypto Session. Thus, Netflix avoids relying solely on HTTPS to protect assets. For instance, from the Android OS view, the app asks the Widevine CDM to decrypt and verify the received messages. The decrypted data are sent back to the app without any particular protection. Therefore, by monitoring \texttt{OEMCrypto\_Generic\_Decrypt}, we were able to intercept all exchanged messages between the CDN and the Netflix app. These messages contain download URLs for \texttt{timedtexttracks} (for subtitles), \texttt{video\_tracks} and \texttt{audio\_tracks}. Each of this category contains multiple links corresponding to different languages for subtitles and audio in addition to different qualities for video. Although the downloaded videos are encrypted using the Widevine CDM, both audio and subtitles can be obtained in clear. We perform several experiments with the audio and subtitles URLs to evaluate their validity. We find that they are accessible from any platform (PC web browsers, smartphones, tablets), by anyone (no need for Netflix account), from anywhere (no location verification), and for a limited period of time (12 hours approximately). We also find that there is no limit of simultaneous accessed devices.

Our learned lesson is that Netflix seemingly makes it harder to spy on messages sent by the CDN by adding an extra layer of protection provided by Widevine. Thus, attackers might be clueless, since defeating Android certificate pinning is not enough. However, WideXtractor easily allowed us to identify the use of the Widevine Generic Crypto API. Thus, it becomes even more straightforward to obtain the exchanged messages in clear by just recovering the returned buffer of \texttt{OEMCrypto\_Generic\_Decrypt}. The advantage of our approach is that we no longer need to bypass certificate pinning implemented by the OTT app. We were surprised that Netflix does not protect audio tracks by a Content Key. During our responsible disclosure, we discovered that Netflix was not even aware of that, because they believed that non-Dash mode was sufficient. We went further and analyzed six other popular OTT apps: Disney+, Amazon Prime Video, Hulu, HBO Max, Starz and Showtime. We find that, unlike Netflix, all of them encrypt their audio tracks with the Content Key.

\subsection{Widevine Over EME}
Similar to the unified DRM API of Android, the World Wide Web Consortium (W3C) defines the Encrypted Media Extensions (EME) standard to provide a standardized API enabling web applications to interact with the browser-supported DRM. EME is designed to make the same web application to run on any browser regardless of the DRM implementation. Despite being optional, EME is supported in major browsers: Edge, Firefox, Chrome, Safari, Opera, and their mobile versions~\cite{eme_browsers}. 

The logic of the EME standard is quite similar to the Android DRM system. Indeed, when the web application attempts to play an encrypted video, it starts by creating a \texttt{MediaKeys}, which is the object providing access to the CDM. Then, it calls \texttt{createSession} to instantiate \texttt{MediaKeySession} managing the lifetime of a DRM license. Next, the \texttt{MediaKeySession} object generates a license request by calling \texttt{generateRequest}. This message is sent to the license server to require the necessary decryption keys. Once the response is received, \texttt{MediaKeySession} calls the \texttt{update} method to parse the obtained license inside the CDM. Now, we can decrypt the media using the keys loaded from the license.

In PC browsers, Widevine comes as a plugin in different browsers, such as Firefox and Chrome, supporting the EME standard. It is true that our work focuses mainly on Android Widevine. In order to overcome this limitation, we study the Widevine flow as it is implemented within the browsers providing EME. Here, we note that the CDM software is obfuscated and hides its symbols. Therefore, we follow a different approach: instead of hooking the browser EME functions, we implemented a browser plugin that intercepts all EME related data. Then, we parse these data and compare them with the ones obtained in Android Widevine. We notice a big reciprocity between the Widevine messages in Android and PC browsers. This confirms that the Widevine protocol in \autoref{fig:protocol} works similarly in different systems. The main difference that we noticed is that the Widevine RoT in browser consists of a whitebox implementation of the \textit{Device RSA Key}.

Thus, we extend WideXtractor to trace the Widevine flow by merely looking into the EME received messages. Our approach has the advantage of successfully following the Widevine flow without regard to the applied obfuscation or the actual called functions. Based on a Chrome EME logger plugin~\cite{emeplugin}, as in WideXtractor we log buffer values and use key usage info from \texttt{update} calls to identify the message purpose within the Widevine protocol. This allowed us to log additional information to link EME calls to Widevine functions.

\section{Discussion} \label{sec:discussion}

Widevine enthusiastically pitches the virtue of their DRM solution. Widevine being proprietary, there is no easy way to verify the security claims of this piece of software running in billions of devices. The goal of our reverse-engineering efforts is to go beyond this market irrationality. In this section, we show how our study conveniently helps in highlighting a gap between what Widevine promises and their technical solution. The raised issues concern not only OTT, but also final users.

\subsection{Privacy Concerns}
The Widevine protocol comes with privacy concerns for users in the streaming ecosystem. These issues are due to the need of Widevine to identify users' devices for bailing purposes. Indeed, Widevine collects device specific data, and sends them to distant servers, such as the provisioning or license ones. For instance, in Android these data includes the Widevine Device ID within the Widevine keybox, and the device blob containing several device-identifying fields, such as the device architecture, phone model, CDM version, or build info.

Ironically, Widevine commits to respect users' privacy. As a matter of fact, Widevine claims to follow the EME standard. Despite being non-normative, user-tracking issues are being pointed out in the privacy section of the EME standard~\cite{eme_browsers}. However, the usage of \textit{Distinctive Identifier} or \textit{Distinctive Permanent Identifier} allows origins crossing information to spot a single user based on these device-unique values. This is harmful for privacy, since it allows third-party servers to profile users' behavior during media consumption. Moreover, users never consent to such device tracking.

\subsection{Recovering Widevine L3 RoT}
Widevine presents their DRM for OTT platforms as a solution to protect them from piracy. There exist several levels of compromise; each one relates to some cryptographic keys in the key ladder. Obviously, RoT recovery constitutes the most severe compromise level, since attackers can derive all keys allowing to decrypt any protected content. Widevine distinguishes L1 RoT and L3 RoT, as it is more challenging to compromise L1 compared to L3. Indeed, Widevine relies on software-only protection mechanisms to hide L3 RoT. It is true that such protection is brittle and doomed to be broken. However, advanced obfuscation techniques might make the compromise quite involving and resources demanding. Here, we show how our understanding of the Widevine protocol may allow attackers to get L3 RoT without specific knowledge of the underlying obfuscation in an automated matter.

As explained in~\autoref{subsec:keybox}, the Widevine RoT is encapsulated inside a keybox that is used to initiate the key ladder in order to retrieve clear content. Starting by certificate provisioning, the RoT is also used in L3 to protect the received \textit{Device RSA Key} for persistent storage (i.e. rewrap operation) or using keybox related data in device blob. Accordingly, we build the following approach to recover L3 RoT. We know that, by design, the RoT must somehow be loaded during the execution of the Widevine protocol, but the applied obfuscation hides the loaded RoT. Here, we rely on WideXtractor to better discern the moment, where the RoT is actually in the memory in clear. At this point, we dynamically analyze all memory regions used during obfuscated cryptographic operations within the Widevine library. We search for the keybox structure (e.g., magic number, device ID). Thus, we were able to recover the L3 keybox on a Nexus 5, including the 128-bit AES \textit{Device Key}, due to an insecure storage of sensitive information (CWE-922). Technical details can be found in appendix~\ref{appendix:weakensses:l3_keybox_recovery}. Our method is efficient, since we limit the spatial and temporal memory monitoring.

\textbf{Responsible Disclosure.} Our findings have been timely reported to all concerned parties following their responsible disclosure process. Netflix was quite responsive and we got rewarded via their bug bounty program. Regarding Google Widevine, our security report was assigned with the highest priority within the Google Vulnerability Reward Program (VRP). The Widevine security team investigated our findings and issued a patch to mitigate our identified flaws. Google assigned the CVE-2021-0639 for us, and acknowledged us in the Google Hall of Fame and the Android Security Acknowledgments. Our goal is to improve the knowledge about DRM, and not to provide copyright infringement tools.

\section{Related Work} \label{sec:related_work}

\subsection{Closed Source Proprietary Protocol}
Closed source protocols are often studied in the literature to provide building grounds or to point out security flaws of the analyzed protocol. For instance, Wouters et al.~\cite{DBLP:journals/tches/WoutersGP21} show that the proprietary autonomous car keyless protocol of Tesla is vulnerable to key injection, which allows an attacker to steal a car in a matter of minutes. Moreover, in their work ARIstoteles, Kröll et al.~\cite{DBLP:conf/esorics/KrollKKHC21} reverse engineered the Apple Remote Invocation undocumented protocol on iOS and found several vulnerabilities. Their work also includes the design of a tool to foster future research on this topic.

\subsection{Widevine Keys Recovery}
In 2019, David Buchanan claimed to have broken L3 Widevine on Linux Chrome browsers in a tweet~\cite{buchananTwitterWidevine19} being the only available information about this attack. Buchanan mentioned that L3 relies on AES-128 whitebox to protect media and was vulnerable to Differential Fault Analysis (DFA). Buchanan has never provided any further detail.

Tomer Hadad released widevine-l3-decryptor on Github at the end of 2020. This project is a Chrome extension on Windows that contains a hard-coded value of an RSA key pair used by Widevine L3. Hadad mentioned that he extracted the RSA private key ``\textit{by applying some mathematical tricks to Arxan's whitebox algorithm}", before releasing a full writeup after Google's patch. Unlike  Buchanan, Hadad explained that the L3 RoT in Chrome browsers is a whitebox of RSA, and not AES. In November 2020, Google issued a DMCA takedown request against widevine-l3-decryptor and all its forks~\cite{google_dmca}, proving that L3 security is still seen by Google as a serious matter. In a BlackHat Asia talk, Zhao~\cite{zhao2021widevine} explained how he broke into Widevine L1 within the TEE to recover the Widevine keybox of a Pixel 4. However, he did not show how a recovered keybox can be used to decrypt protected contents. In our work, we took this further step and implemented the cryptographic mechanisms of Widevine.

\section{Conclusion} \label{sec:conclusion}

In this paper, we presented the undocumented closed-source Widevine protocol with its cryptographic components. By reverse engineering the Widevine CDM on Android, we extracted the logic behind its key ladder and provisioning phases. Based on the gained insights, we design WideXtractor, a tool analyzing the protocol workflow and all message exchanges between clients and distant servers. We show the effectiveness of WideXtractor by inspecting the use of Widevine by Netflix, thereby uncovering interesting findings about Netflix asset protection. Furthermore, we were able to trivially recover the L3 RoT, which allows attackers to obtain any content of sub-HD quality. Being widely deployed, DRM security becomes critical. Our objective is to encourage and foster further research about DRM-related technologies.

\bibliographystyle{IEEEtran}
\bibliography{references}

\newpage

\appendices

\section{OEM Crypto Library Symbols Equivalents} \label{appendix:oeccSymbol}
\begin{table}[H]
\center
\begin{adjustbox}{max width=0.45\textwidth}
\begin{threeparttable}
\begin{tabular}{|c|l|c|l|}
\hline
\textbf{Symbols} & \textbf{OEMCrypto Functions} & \textbf{Symbols} & \textbf{OEMCrypto Functions} \\ \hline
\textbf{oecc01} & Initialize & \textbf{oecc25} & Generic\_Decrypt \\ \hline
\textbf{oecc02} & Terminate & \textbf{oecc26} & Generic\_Sign \\ \hline
\textbf{oecc03} & InstallKeybox & \textbf{oecc27} & Generic\_Verify \\ \hline
\textbf{oecc04} & GetKeyData & \textbf{oecc28} & GetHDCPCapability \\ \hline
\textbf{oecc05} & IsKeyboxValid & \textbf{oecc29} & SupportsUsageTable \\ \hline
\textbf{oecc06} & GetRandom & \textbf{oecc30} & UpdateUsageTable \\ \hline
\textbf{oecc07} & GetDeviceID & \textbf{oecc31} & DeactivateUsageEntry \\ \hline
\textbf{oecc08} & WrapKeybox & \textbf{oecc32} & ReportUsage \\ \hline
\textbf{oecc09} & OpenSession & \textbf{oecc33} & DeleteUsageEntry \\ \hline
\textbf{oecc10} & CloseSession & \textbf{oecc34} & DeleteUsageTable \\ \hline
\textbf{oecc11} & DecryptCTR & \textbf{oecc35} & LoadKeys* \\ \hline
\textbf{oecc12} & GenerateDerivedKeys & \textbf{oecc36} & GenerateRSASignature* \\ \hline
\textbf{oecc13} & GenerateSignature & \textbf{oecc37} & GetMaxNumberOfSessions \\ \hline
\textbf{oecc14} & GenerateNonce & \textbf{oecc38} & GetNumberofOpenSessions \\ \hline
\textbf{oecc15} & LoadKeys* & \textbf{oecc39} & isAntiRollbackHwPresent \\ \hline
\textbf{oecc16} & RefreshKeys & \textbf{oecc40} & CopyBuffer \\ \hline
\textbf{oecc17} & SelectKey* & \textbf{oecc41} & QueryKeyControl \\ \hline
\textbf{oecc18} & RewrapDeviceRSAKey & \textbf{oecc42} & LoadTestKeybox \\ \hline
\textbf{oecc19} & LoadDeviceRSAKey & \textbf{oecc43} & ForceDeleteUsageEntry \\ \hline
\textbf{oecc20} & GenerateRSASignature* & \textbf{oecc44} & GetHDCPCapability \\ \hline
\textbf{oecc21} & DeriveKeysFromSessionKey & \textbf{oecc45} & LoadTestRSAKey \\ \hline
\textbf{oecc22} & APIVersion & \textbf{oecc46} & Security\_Patch\_Level \\ \hline
\textbf{oecc23} & GetSecurityLevel	& \textbf{oecc47} & LoadKeys* \\ \hline
\textbf{oecc24} & Generic\_Encrypt & \textbf{oecc48} & DecryptCENC \\ 
\hline

\end{tabular}
		\begin{tablenotes}
        \footnotesize
        \item[*] Duplicated entries differ in version.
    \end{tablenotes}
\end{threeparttable}
\end{adjustbox}
\end{table}

\section{L3 Keybox Recovery} \label{appendix:weakensses:l3_keybox_recovery}

Being the root of trust, we are motivated to recover the keybox. Widevine maintains a different keybox for the different levels of security. In~\autoref{subsec:keybox}, we explained that L1 protection is TEE-dependent. In QSEE, it is based on the Secure File System, whose security is outside the scope of this paper. Here, we will focus on L3 keybox. Note that L3 implementations are diverse. Our analysis shows that Widevine is as secure as the weakest one, since license keys for a given media are shared among all L3 implementations. Therefore, someone might take advantage of outdated implementations to break into Widevine. Indeed, they can intentionally display content on vulnerable smartphones, so that they can easily recover protected media. This works as long as OTT platforms keep support for old Android smartphones, as they target wide audience. In this paper, we study the L3 of Google Nexus 5 that still runs many OTT apps.

By taking a closer look at \texttt{libwvdrmengine.so}, we notice that \texttt{OEMCrypto} L3 functions are obfuscated. This makes our analysis more complex, since the keybox is only used within these functions. Moreover, we find that all obfuscated functions apply anti-reverse transformations, such as control flow flattening, that make static analysis less relevant. In addition, memory regions are mapped with read and execute permissions. Because of ARM architecture blurring line between code and data, we find it hard to tell if these mapped regions are destined for data to load or code to execute.

The approach that we followed to recover the keybox was not to directly break into the layer of obfuscation. This would have made of our work technology-dependent, while we aim for more long-term lessons. Instead, we stepped back and monitored the unprotected functions calling the \texttt{OEMCrypto} interface using WideXtractor. Indeed, we notice that most functions of \texttt{libwvdrmengine.so} are not protected. Thus, we managed to collect a lot of memory data loaded during the execution of the obfuscated functions. Of particular interest, we were able to observe all memory unmapping that happens through calls to \texttt{munmap}. We noticed that the \texttt{OEMCrypto} functions load a significant amount of data though these calls especially sensitive ones.

Thus, our next target is to look for a function requiring the keybox for its operations. We recall that the keybox is used in \texttt{OEMCrypto\_RewrapDeviceRSAKey} to re-encrypt the provisioned Device RSA Key, but also during device blob creation with \texttt{OEMCrypto} methods like \texttt{GetDeviceID} or \texttt{GetKeyData}. Accordingly, these functions map a proper region of memory for the keybox, loads the keybox value inside it, and finally unmaps that region at the end of the function. Because of the obfuscation, it is hard to observe the loading step. However, these functions do not clear the memory before unmapping. Therefore, we retrieved the content of the unmapped regions. Then, relying on what we know about the keybox, we filtered this content to keep the regions of size 128 bytes including the keybox magic number.

It turns out that there is only one. We verify our finding by checking the CRC-32 value. The keybox being recovered, we can now decrypt the license keys, hence the video tracks destined to L3. It is worth noting that this is particularly interesting, especially that we did not even get to break into the underlying obfuscation. In fact, our analyses were guided by the conceptual structure of the Widevine protocol.

\end{document}